# Direct measurement of acoustic spectral density and fractional topological charge

Hao Ge[1, §], Zi-Wei Long[1, §], Xiang-Yuan Xu[1], Yang Liu[2,3], Bi-Ye Xie[4], Jian-Hua Jiang[2,3, *], Ming-Hui Lu[1,5, †], and Yan-Feng Chen[1, 5, ‡]

[1] *National Laboratory of Solid State Microstructures & Department of Materials Science and Engineering, Nanjing University, Nanjing, Jiangsu 210093, China*

[2] *Institute of Theoretical and Applied Physics, School of Physical Science and Technology, & Collaborative Innovation Center of Suzhou Nano Science and Technology, Soochow University, 1 Shizi Street, Suzhou, 215006, China*

[3] *Key Lab of Advanced Optical Manufacturing Technologies of Jiangsu Province & Key Lab of Modern Optical Technologies of Ministry of Education, Soochow University, Suzhou 215006, China.*

[4] *School of Science and Engineering, The Chinese University of Hong Kong, Shenzhen, 518172, China.*

[5] *Jiangsu Key Laboratory of Artificial Functional Materials, Nanjing, Jiangsu 210093, China.*

[§] *These authors contributed equally to this work.*

*Correspondence should be addressed to: joejhjiang@hotmail.com, luminghui@nju.edu.cn or yfchen@nju.edu.cn*

**Abstract:**

Local density-of-states (LDOS) is a fundamental spectral property that plays a central role in various physical phenomena such as wave–matter interactions. Here, we report on the direct measurement of the LDOS of acoustic systems and derive from which the fractional topological number in an acoustic Su-Schrieffer-Heeger system. The acoustic LDOS is quantified here with a state-of-the-art technique through the measurement of the volume flow rate and the acoustic pressure with a local excitation-probe configuration. Based on this method, we study the acoustic Purcell effect and establish experimentally the important relation between the near-field LDOS and the far-field acoustic emission power. Moreover, we detect the LDOS in the one-dimensional acoustic Su-Schrieffer-Heeger model and observe the fractional topological number of the system. Our work unveils the important role of the LDOS in acoustic phenomena and paves the way toward characterizing and tailoring the LDOS in topological systems.

*Introduction.*

The bulk density-of-states (DOS) gives the spectral distribution of the eigenstates of a system which, in condensed matter context, is closely related to the optical, thermal, and transport properties of the underlying material. In quantum electrodynamics, the Purcell effect reveals the photonic DOS as one of the central quantities that determines the spontaneous emission rate of a quantum emitter [1-10]. However, the importance of DOS in acoustics and its relevance to many acoustic phenomena were uncovered only in recent years. With the theory of acoustic Purcell effect [1], it was found that the LDOS of acoustic waves determines the sound emission rate of a subwavelength acoustic source [11-17]. However, the direct measurement of acoustic LDOS has not yet been achieved, mainly due to the technical challenges in characterizing the acoustic properties of subwavelength monopole sources.

In this work, we develop a method to directly measure the acoustic LDOS via the integration of an acoustic particle velocity sensor and a micro-electro-mechanical system (MEMS) pressure microphone. Theoretically, the acoustic LDOS is related to the imaginary part of the onsite acoustic Green's function. From the acoustic wave equations, the LDOS can be extracted from the frequency and spatial dependence of the volume flow rate and the acoustic pressure when the acoustic waves are excited by a subwavelength source. We demonstrate the underlying principle by measuring the acoustic LDOS of two prototypes of resonators: the Helmholtz resonator and Fabry-Pérot resonator. In these experimental systems, we verify the acoustic Purcell effect and reveal the important relation between the far-field acoustic emission power and the near-field acoustic LDOS. Furthermore, we show that acoustic LDOS can be used to directly extract the fractional topological number in the celebrated SSH model when it is realized in acoustic metamaterials [18-21], revealing a fundamental phenomenon in topological acoustics.

*Green's function and acoustic LDOS.*

The acoustic LDOS is related to the imaginary part of the onsite Green's function as follows [22]:

$$\rho(\omega, \boldsymbol{r}) = \mp \frac{1}{\pi} \frac{dk_0^2(\omega)}{d\omega} \mathrm{Im} G^{\pm}(\omega, \boldsymbol{r} = \boldsymbol{r}') \tag{1}$$

where $\boldsymbol{r}'$ and $\boldsymbol{r}$ denote the positions of the source and the detector, respectively, $G^{\pm}$ is the retarded/advanced Green's function, and $k_0^2 = \omega^2/c_0^2$ ($c_0$ is the speed of sound). Assuming the Sommerfeld radiation boundary condition, the retarded Green's function is the solution to the acoustic wave equation with a monopole source of unit strength:

$$\nabla^2 G^+(\omega, \boldsymbol{r}, \boldsymbol{r}') + k_0^2 G^+(\omega, \boldsymbol{r}, \boldsymbol{r}') = \delta(\boldsymbol{r} - \boldsymbol{r}') \tag{2}$$

The corresponding equation for the acoustic pressure $p$ is:

$$\nabla^2 p + k_0^2 p = i\omega \rho_{air} u_s S_s \delta(\boldsymbol{r} - \boldsymbol{r}') \tag{3}$$

where $\rho_{air}$ is the mass density of air, $u_s$ is the acoustic particle velocity on the surface of monopole source, and $S_s$ is the surface area of the source. The product of $u_s$ and $S_s$ is the volume flow rate that represents the strength of the source. By comparing equation (2) and (3), the Green's function at the source location can be expressed as:

$$G^+(\omega, \boldsymbol{r} = \boldsymbol{r}') = \frac{p_s}{i\omega \rho_{air} u_s S_s} \tag{4}$$

where $p_s$ denotes the acoustic pressure at the source location. The acoustic LDOS is then obtained as:

$$\rho(\omega, \boldsymbol{r}) = \frac{2}{\pi} \frac{1}{\rho_{air} c_{air}^2} \text{Re}\left(\frac{p_s}{u_s S_s}\right). \tag{5}$$

Therefore, the acoustic LDOS at the source location is proportional to the radiation resistance of the sound source.

*Methods for measuring the acoustic LDOS.*

From equation (5), it can be seen that the acoustic LDOS at the source location can be obtained via the acoustic pressure $p_s$ and the volume flow rate of source $u_s S_s$. However, the direct measurement of $u_s S_s$ of a small-size monopole source is quite challenging. Here, we design a compact detector tube which combines a balanced armature speaker (Knowles, CI-22955-000) as the sound emitter, a homemade acoustic particle velocity sensor, and a Knowles MEMS silicon microphone, as shown in Fig. 1(a). The detector tube is connected to the structure and the sensors are placed on the contact surface. The acoustic pressure $p_s$ and velocity $u_s$ at the source location are measured, respectively, by the MEMS pressure microphone and the particle velocity sensor. In our setup, $S_s$ becomes the cross-sectional area of the detector tube. In our experiment, the detector tube has a width of 6 mm and a height of 10 mm [Fig. 1(b)] which are of deep subwavelength scales. This design gives an excitation-probe setup with a point-like source and a local detector which is targeted to probe the Green's function in Eq. (4).

In an acoustic cavity such as the Helmholtz resonator [Fig. 1(a)], the resonant LDOS is known to be of the form [23]:

$$\rho(\omega_n) = \frac{2}{\pi \omega_n} \frac{Q}{V}, \tag{6}$$

where $\omega_n$ is the resonance frequency, $Q$ is the quality factor, and $V$ is the modal volume. For the lowest resonant mode, its wavefunction is nearly homogeneous inside the cavity. The acoustic waves radiate out into the surrounding air through the neck. For a cubic Helmholtz resonator with a side width of 60 mm, we calculate the acoustic LDOS at the resonant frequency for different neck sizes. As shown in Fig. 1(c), the results from Eq. (6) agree well with the results from finite-element simulations based on Eq. (5), which demonstrate the feasibility of determining the acoustic LDOS via Eq. (5).

*Acoustic Purcell effect.*

We now reveal the acoustic Purcell effect through setups with two different cavities: the Helmholtz resonator [Fig. 2(a)] and the Fabry-Pérot resonator [Fig. 2(c)]. To avoid the subtle influence of residue and environment waves, the experiments are conducted in an anechoic chamber. The speaker is driven by the sound card and the signals of sensors are acquired by the NI 9234 data acquisition module. As shown in Figs. 2(b) and 2(d), the measured data are in good agreement with the simulation results. The overall LDOS of the Helmholtz resonator is higher than the Fabry-Pérot resonator, which is due to the stronger wave confinement in the Helmholtz resonator. The spatial distribution of LDOS is proportional to the squared acoustic pressure field inside the cavity: $\rho(\omega, \boldsymbol{r}) \propto |p(\omega, \boldsymbol{r})|^2$. For the Helmholtz resonator, the LDOS is nearly uniform in the cavity. In contrast, for the Fabry-Pérot resonator, the LDOS is inhomogeneous and reaches its maximum at the bottom of the cavity where we probe the LDOS.

We now demonstrate that the far-field sound emission power is proportional to the acoustic LDOS at the source location. For this purpose, two Helmholtz resonators with different side-widths (60 mm and 63 mm, separately) are used. To probe the far-field radiation power, which is proportional to the square of the acoustic pressure $|p|^2$, we use a microphone 30 cm away from the resonator to detect the radiation [Fig. 3(a)]. The measured LDOS and the detected (normalized) far-field acoustic radiation intensity $|p|^2$ are shown together in Figs. 3(b) and 3(c) for the two resonators. For both systems, the excellent proportionality between the LDOS and the far-field radiation confirms in experiments the acoustic Purcell effect which gives a remarkable connection between the near- and far-field acoustic properties.

*Acoustic SSH model and fractional topological number.*

The LDOS is also tightly connected to the fractional quantum number in the SSH model and other topological systems [24-28]. Here, we use the acoustic SSH model as a prototype to measure the fractional topological number via the LDOS using the above method. Here, the 1D acoustic SSH model is realized by coupling an array of acoustic cavities in a dimerized way [Fig. 4(a)]. Each cavity has a height of $L_0 = 150$ mm and a side width of $D_0 = 30$ mm. Two types of square tubes of different widths (5 mm and 12 mm, respectively) are used to connect the cavities and introduce the dimerized couplings. All tubes are of length 60 mm. When the intra-cell coupling is stronger (weaker) than the inter-cell coupling, the system is trivial (topological). Figs. 4(b) and 4(c) show the acoustic eigenstate spectrum from finite-element simulations for the topological and trivial cases. There are in total 16 eigenstates in the spectrum for both cases, which equals to the number of cavities in our system. For the topological case, there are two edge states emerging at the middle of the bandgap.

The sample is fabricated using 3D printing technology [Fig. 4(d)] and has eight unit-cells. There are two sites within each unit cell. In the experiment, the detector tubes are placed at the top of the cavities. We measure the LDOS of each cavity with a frequency step of 2 Hz. We define that the LDOS at the $i$-th site is $\rho(f,i) = \frac{1}{2}\rho_{max}(f,i)L_0 D_0^2$ where $\rho_{max}(f,i)$ is the maximum LDOS in the $i$-th cavity (i.e., the LDOS measured at the top of the cavity, see Supplemental Material for details [29]).

In Fig. 4(e), we present the LDOS $\rho(f,i)$ for the first four sites of the system using results from theoretical calculations, finite-element simulations, and the experimental data (see Supplemental Material for details [29]) for the topological case. These results are in agreement with each other. We find that for the LDOS in site 1, the edge state emerges as a single peak in the band gap. For the LDOS in sites 2-4, the two peaks are associated with the bulk states. From these results, we can infer that the edge states are mainly localized in the edge cavities.

The SSH model supports the fractional topological number at edges which can be understood via the concept of filling anomaly [30, 31]. In analog of the fermionic band filling of the valence band and the edge states, here the mode charge for each site is defined as:

$$q_i = \int_0^{f_{gap}} \rho(f,i)\,df. \qquad (7)$$

Here, $f_{gap}$ is a frequency in the band gap and is above the edge states. The mode charge of each unit cell is the sum of the mode charge of the two sites. The results are shown in Fig. 4(f). For the trivial case, the mode charge of each unit cell is close to the theoretical value of 1. For the topological case, it is shown that the edge unit cell has a mode charge close to 1.5, which reveals the 1/2 fractional quantum number for the topological SSH model.

*Conclusion and outlook.*

In this work, we first verify experimentally the acoustic Purcell effect that provides an important relation between the near-field LDOS and the far-field acoustic emission power. We then observe in an acoustic SSH system the emergence of 1/2 fractional topological number which is a celebrated prediction of the Jackiw-Rebbi theory. These findings reveal the central importance of the LDOS in acoustic systems and thus opens a pathway toward the fundamental physics of acoustic phenomena [21,32].

*Acknowledgments*.
The work is jointly supported by the National Key R&D Program of China (Grant Nos. 2021YFB3801801, 2017YFA0303702, 2017YFA0305100 and 2018YFA0306200), the National Natural Science Foundation of China (Grant Nos. 11890702, 12125504, 51721001 and 12074281). We also acknowledge the support of the Natural Science Foundation of Jiangsu Province, Jiangsu specially-appointed professor funding, and the support from the Academic Program Development of Jiangsu Higher Education

(PAPD).

**References**.

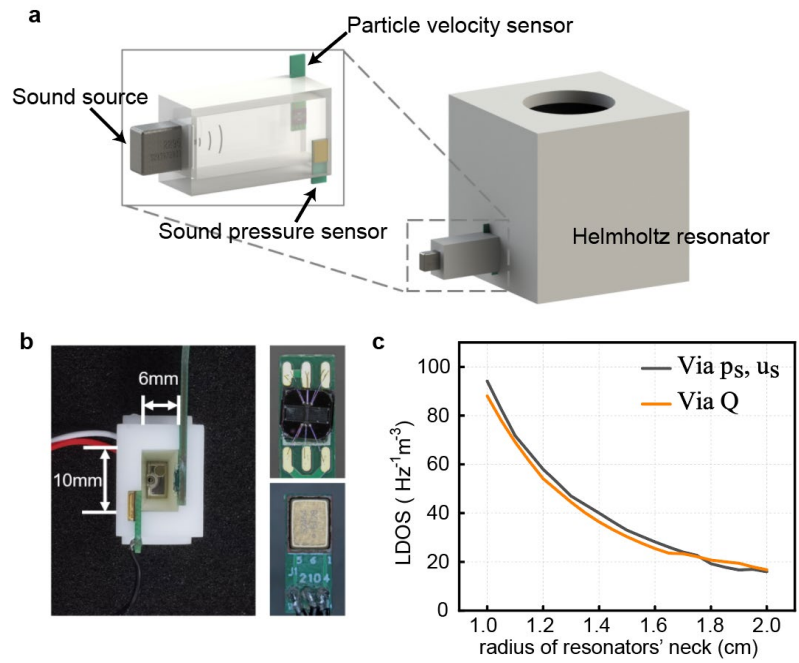

**Fig. 1** (a) A compact detector tube which combines a balanced armature speaker, a homemade acoustic particle velocity sensor, and a Knowles MEMS silicon microphone. The tube is connected to the testing structure, and can be considered as a subwavelength source. The acoustic pressure and the volume flow rate of source are measured by the sensors. The acoustic LDOS at the source location then can be extracted. (b) Photos of the detector tube, velocity sensor and the MEMS microphone. The cross-sectional area of the tube is 10 mm×6 mm. (c) The simulation results of acoustic LDOS at the resonance frequency for a Helmholtz resonator with different neck sizes. The results from the quality factor $Q$ of the resonator agree well with the results from the acoustic pressure and velocity at the source location.

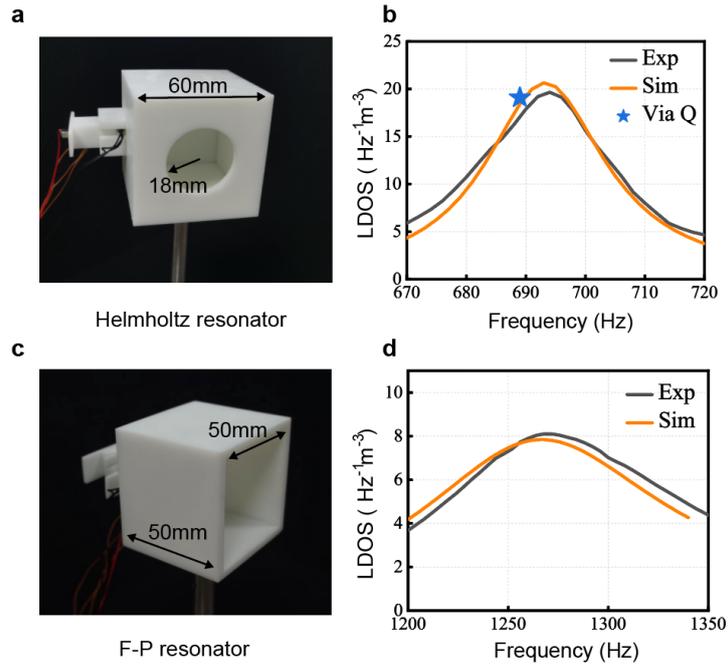

**Fig. 2** (a) Photo of the Helmholtz resonator. The detector tube is connected to the resonator and the sensors are placed on the contact surface. (b) The experimental and simulation results of the acoustic LDOS spectrum for the Helmholtz resonator. The blue star corresponds to the resonant LDOS calculated by the quality factor of the resonator. (c) Photo of the Fabry-Pérot resonator. The detector tube is placed at the bottom of the resonator, where the LDOS reaches its maximum. (d) The experimental and simulation results of the acoustic LDOS spectrum for the Fabry-Pérot resonator.

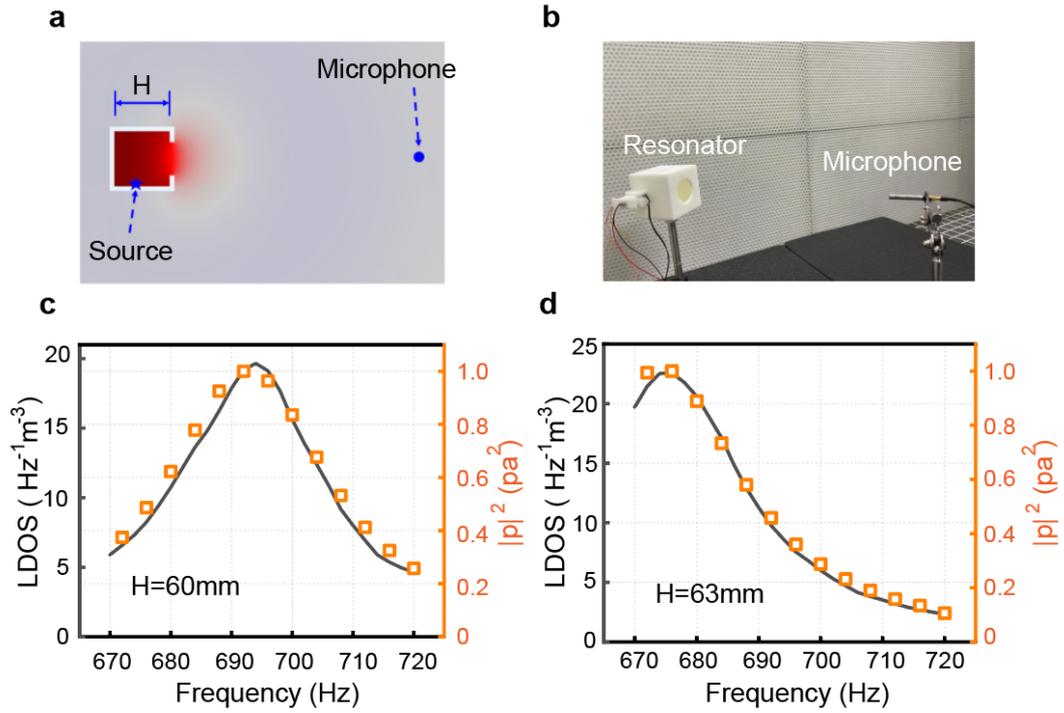

**Fig. 3** (a), (b) Experimental setup. The microphone is placed 30 cm away from the resonator to detect the radiation field. The detector tube is connected to the resonator to emit the sound and probe the LDOS at the source location. (c) The measured LDOS spectrum (black curve) and the normalized far-field radiation $|p|^2$ (orange square) for the Helmholtz resonator with a height of 60 mm. (d) The measured LDOS spectrum and the normalized far-field radiation $|p|^2$ for the Helmholtz resonator with a height of 63 mm.

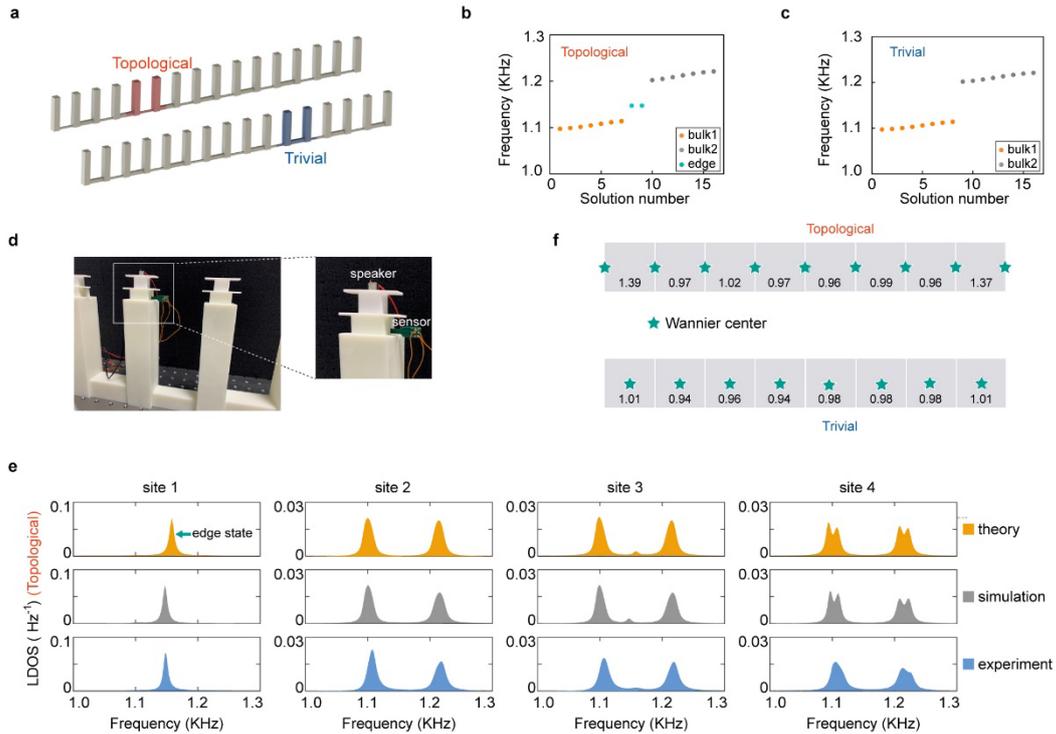

**Fig. 4** (a) The structure of the 1D acoustic SSH model. The eigenstate spectrum from finite-element simulations for the topological (b) and trivial case (c). A pair of edge states emerge in the bandgap for the topological case. Here, "bulk 1" and "bulk 2" label the valence and conduction bands, respectively. (d) Photo of the sample. Inset shows the structure of the device for the probe of the acoustic LDOS. (e) The LDOS spectrum of the first four sites of the system. In the LDOS spectrum of site 1, the edge state emerges as a single peak in the band gap. In the LDOS spectrum of sites 2-4, the two peaks correspond to the bulk states. (f) Measured mode charge of each unit cell for the topological and trivial case. The green stars label the Wannier centers.